\documentclass[reprint,nofootinbib,superscriptaddress,amsmath,amssymb,aps,prd]{revtex4-1}
\usepackage{amsmath,mathrsfs,amsbsy,color,graphicx,bm,amsthm,amsfonts}
\usepackage{bbm}
\usepackage{subcaption}
\usepackage{times}
\usepackage{units}
\usepackage{dcolumn}
\usepackage{graphicx}
\usepackage{epsfig}
\usepackage{epstopdf}
\usepackage[colorlinks,linkcolor=red,anchorcolor=green,citecolor=blue,CJKbookmarks=True]{hyperref}
\DeclareMathSymbol{\shortminus}{\mathbin}{AMSa}{"39}
\usepackage{mathrsfs}
\usepackage{braket}
\usepackage{amssymb}
\usepackage{txfonts}
\usepackage{float}
\usepackage{enumitem}
\usepackage{multirow}

\begin{document}
	
	\title{Gravitational waveforms from periodic orbits around a charged black hole with scalar hair}
	
	\author{Weike Deng}
	\affiliation{School of Science, Hunan Institute of Technology, Hengyang 421002, China}
	\affiliation{Department of Physics, Key Laboratory of Low Dimensional Quantum Structures and Quantum Control of Ministry of Education, and Synergetic Innovation Center for Quantum Effects and Applications, Hunan Normal
	University, Changsha  410081, China}
	\affiliation{Hengyang Vocational College of Science and Technology, Hengyang 421001, China}

	\author{Sheng Long}
	\affiliation{School of Fundamental Physics and Mathematical Sciences, Hangzhou Institute for Advanced Study, University of Chinese Academy of Sciences, Hangzhou 310024, China}
	
	\author{Qin Tan}
\email[]{tanqin@hunnu.edu.cn (Corresponding authors)} 
\affiliation{Department of Physics, Key Laboratory of Low Dimensional Quantum Structures and Quantum Control of Ministry of Education, and Synergetic Innovation Center for Quantum Effects and Applications, Hunan Normal
	University, Changsha  410081, China}	
	
	\author{Jiliang Jing}
	\email[]{jljing@hunnu.edu.cn (Corresponding authors)} 
	\affiliation{Department of Physics, Key Laboratory of Low Dimensional Quantum Structures and Quantum Control of Ministry of Education, and Synergetic Innovation Center for Quantum Effects and Applications, Hunan Normal
		University, Changsha 410081, China}

\begin{abstract}

We investigate geodesic motion and gravitational-wave signatures of charged black holes with scalar hair. Using the effective potential approach, we analyze marginally bound orbits and innermost stable circular orbits, showing how their positions and energy thresholds are modified by the scalar hair parameter $r_B$. These results demonstrate scalar hair's role in altering the boundary of stable motion. We further explore periodic orbits characterized by rational frequency ratios, labeled by the index $(z,w,v)$, and quantify how scalar hair affects their orbital energy and angular momentum. Based on these orbital properties, we compute gravitational waveforms from extreme mass-ratio inspirals where a stellar-mass compact object orbits a supermassive charged black hole with scalar hair. Using the numerical kludge method, we generate waveforms that exhibit clear zoom-whirl patterns with morphology visibly affected by $r_B$. Our results show that scalar hair leaves distinguishable imprints on waveforms, suggesting future space-based detectors could probe deviations from classical black hole spacetimes through extreme mass-ratio inspirals observations.

\end{abstract}
	
	\maketitle
	\section{Introduction}

Black holes represent one of the most remarkable predictions of Einstein's general relativity, serving as natural laboratories for testing strong-field gravity. The advent of gravitational-wave astronomy has opened new observational windows into these compact objects. Ground-based detectors like LIGO and Virgo have confirmed stellar-mass black hole coalescences, providing strong evidence for general relativity in the dynamical, strong-field regime \cite{BP2016,BP20161,BP20162,BP20163,LIGOScientific:2017bnn,LIGOScientific:2017vox,LIGOScientific:2017ycc,LIGOScientific:2018mvr}. Future space-based missions including LISA \cite{LISA}, Taiji \cite{TJ}, TianQin \cite{TQ}, and DECIGO \cite{Deligo} will extend this frontier by targeting extreme mass-ratio inspirals (EMRIs), where stellar-mass compact objects spiral into supermassive black holes \cite{EMRI1}. These systems are particularly valuable because the small companion traces intricate geodesics, mapping the central object's spacetime with exquisite precision. Consequently, EMRIs offer unique probes of black hole structure and potential deviations from general relativity \cite{EMRI2,EMRI3,LISA:2022yao}.

Orbital dynamics around black holes play crucial roles in both theoretical studies and astrophysical applications \cite{Dy1,Dy2,Dy3,Dy4,Tan:2024hzw}. Among these, marginally bound orbits (MBOs) and innermost stable circular orbits (ISCOs) are particularly important: MBOs separate bound from unbound motion, while ISCOs define accretion disk inner edges and serve as key parameters in gravitational-wave modeling \cite{B1,B2}. Beyond these fundamental orbits, periodic trajectories characterized by rational frequency ratios exhibit zoom-whirl dynamics, where particles alternate between near-circular whirls and extended excursions \cite{Levin,Dy1,Dy2,Albanesi:2021rby,Gold:2009hr}. Such orbits illuminate phase-space structure, connecting regular circular motion with chaotic scattering, and leave distinctive imprints on gravitational waveforms. The periodic orbit framework has been applied to various black hole spacetimes, including charged black holes \cite{P1}, naked singularities \cite{P2}, Einstein-Lovelock black holes \cite{P4}, Kehagias-Sfetsos solutions \cite{P5}, quantum-corrected black holes \cite{P6}, and brane-world black hole \cite{P7}, with further developments in Refs.~\cite{P8,P9,P10,P11,P12,P13,P14,P15,P16,P17,Junior:2024tmi,Gong:2025mne,Wang:2025hla}. Collectively, studying MBOs, ISCOs, and periodic orbits provides a powerful framework for probing black hole geometries and testing general relativity \cite{B2}.

Recently, Bah et al. introduced five-dimensional topological stars/black holes within Einstein-Maxwell theory \cite{Bah2021,Bah2022}. This construction yields spacetimes smooth at microstate geometry levels while reproducing classical black hole properties macroscopically, enabling exploration of observational signatures. Previous work investigated charged particle dynamics in this background \cite{Stotyn:2011tv,YK2021}. Through Kaluza-Klein dimensional reduction, the five-dimensional Einstein-Maxwell theory maps to an effective four-dimensional Einstein-Maxwell-Dilaton theory admitting static, spherically symmetric solutions \cite{Bena2020,Bena2021}. These solutions enable studies of observable phenomena including gravitational-wave emission, black hole shadows \cite{shadow,Chen:2024aaa,Liu:2024lbi,Zhang:2024jrw,Chen:2023wzv,Zare:2024dtf}, and quasinormal modes \cite{QNM,Guo:2023vmc,Liu:2023uft}, offering deeper insights into charged black hole properties.


This work systematically investigates orbital dynamics in these backgrounds \cite{B1}, including MBOs, ISCOs \cite{B2}, and periodic orbit families characterized by rational frequency ratios. To connect theory with observation, we employ the numerical kludge approximation for computing EMRI waveforms around these black holes \cite{klu,Jing:2025utt,Qiao:2024gfb,Deng:2024ayh,Long:2024axi,Chen:2023lsa,Jing:2023okh,Jing:2023vzq,Delos:2024poq}. This method provides efficient waveform modeling: while not capturing full relativistic radiation structure, it reproduces main spectral features associated with orbital motion, including zoom-whirl dynamics. It thus serves as a useful tool for exploring how background geometry modifications, such as scalar hair presence, affect waveform morphology. We apply this method to generate representative signals from inspirals into charged black holes with scalar hair and analyze their spectral content. 

The paper is organized as follows. Section~\ref{sec2} overviews the charged black hole with scalar hair and derives the effective potential. Section~\ref{sec3} analyzes MBO and ISCO properties and their modification by $r_B$. Section~\ref{sec4} studies the rational number $q$ for periodic orbits and $r_B$'s effects on their energy and angular momentum. Section~\ref{sec5} presents gravitational waveforms from test objects on periodic orbits and their spectral properties. Conclusions and discussions appear in Section~\ref{sec6}. Throughout, we use geometrized units with $G = c = 1$.

\section{The charged black hole with scalar hair}\label{sec2}
We briefly review geodesic orbits around charged black holes with scalar hair. Starting from five-dimensional Einstein-Maxwell theory, the spherically symmetric metric ansatz is \cite{Stotyn:2011tv}
\begin{align}
	\nonumber 
	ds^2 = & -f_S(r)dt^2 + f_B(r)dy^2 + \frac{1}{f_S(r)f_B(r)}dr^2 \\
	& + r^2d\theta^2 + r^2\sin^2\theta d\phi^2,\label{metric1}
\end{align}
with
\begin{align}
	f_B(r) &= 1 - \frac{r_B}{r}, \\
	\quad f_S(r) &= 1 - \frac{r_S}{r}.
\end{align}
This geometry exhibits two coordinate singularities: at $r = r_S$ (event horizon) and $r = r_B$ (where the $y$-circle shrinks to zero size, signaling spacetime termination). Bah et al. showed that after coordinate redefinitions, $r = r_B$ actually hosts a smooth bubble \cite{Bena2020,Bena2021}. For $r_S \geq r_B$, the bubble lies inside the horizon, forming a black string configuration. For $r_S < r_B$, no horizon is encountered since spacetime caps off smoothly at the bubble, interpreted as a topological star \cite{Bena2020,Bena2021}. The metric~\eqref{metric1} rewrites as
\begin{align}
	ds_5^2 = & e^{2\Phi}ds_4^2 + e^{-4\Phi}dy^2, \\
	ds_4^2 = & f_B^{\frac{1}{2}}\left(-f_Sdt^2 + \frac{dr^2}{f_Bf_S} + r^2d\theta^2 + r^2\sin^2\theta d\phi^2\right),
\end{align}
where $\Phi$ is a dilaton field with $e^{2\Phi} = f_B^{-1/2}$. We focus on the $r_B < r_S$ case of the four-dimensional metric, corresponding to charged black holes with scalar hair.

We now study massive test particle orbits around this black hole. The test particle Lagrangian is \cite{B1}
\begin{align}\label{eq1}
	\mathscr{L} = \frac{m}{2}g_{\mu\nu}\frac{dx^\mu}{d\tau}\frac{dx^\nu}{d\tau},
\end{align}
where $\tau$ and $m$ denote proper time and test particle mass, respectively. We set $m=1$ for simplicity. The generalized momentum is
\begin{align}
	p_\mu = \frac{\partial\mathscr{L}}{\partial\dot{x}^\mu} = g_{\mu\nu}\dot{x}^\nu,
\end{align}
where overdots indicate proper time derivatives. The equations of motion are
\begin{align}\label{eqmotion}
	\nonumber
	p_t & = f_B(r)^{\frac{1}{2}}f_S(r)\dot{t} = E, \\
	 p_r &= \frac{\dot{r}}{f_B(r)^{\frac{1}{2}}f_S(r)}, \\
	p_\theta & = f_B(r)^{\frac{1}{2}}r^2\dot{\theta}, \\
	p_\phi &= f_B(r)^{\frac{1}{2}}r^2\sin^2\theta\dot{\phi} = L,
\end{align}
where $E$ and $L$ represent energy and orbital angular momentum per unit mass, respectively. This yields
\begin{eqnarray}
	\dot{t} & = \frac{E}{f_B(r)^{\frac{1}{2}}f_S(r)}\label{eqt}, \\
	\dot{\phi} & = \frac{L}{f_B(r)^{\frac{1}{2}}r^2\sin^2\theta}\label{eqphi}.
\end{eqnarray}

Restricting to spherical backgrounds, we assume equatorial orbits with $\theta = \pi/2$ and $\dot{\theta} = 0$. The Lagrangian~\eqref{eq1} becomes
\begin{align}
	-\frac{E^2}{f_B(r)^{\frac{1}{2}}f_S(r)} + \frac{L^2}{f_B(r)^{\frac{1}{2}}r^2} + \frac{\dot{r}^2}{f_B(r)^{\frac{1}{2}}f_S(r)} = -1,
\end{align}
and the radial effective potential is
\begin{align}\label{eq2}
	V_{\mathrm{eff}} = \left(\sqrt{f_B(r) + \frac{L^2}{r^2}}\right)f_S(r).
\end{align}

\begin{figure}
	\begin{subfigure}{0.7\linewidth}
		\includegraphics[width=\linewidth]{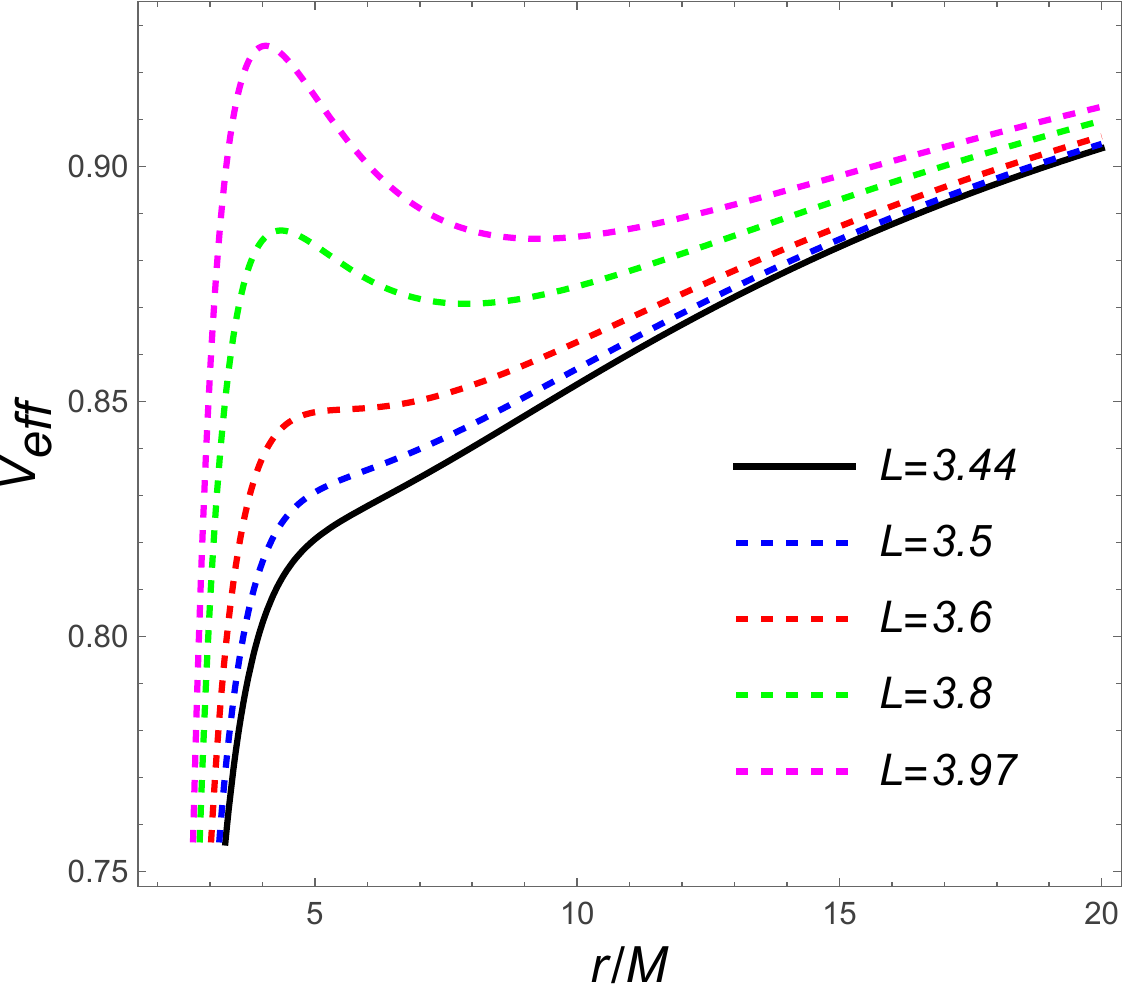}
		\caption{$r_B=1$}
		\label{Fig1a}
	\end{subfigure}\\[6pt]
	\begin{subfigure}{0.7\linewidth}
		\includegraphics[width=\linewidth]{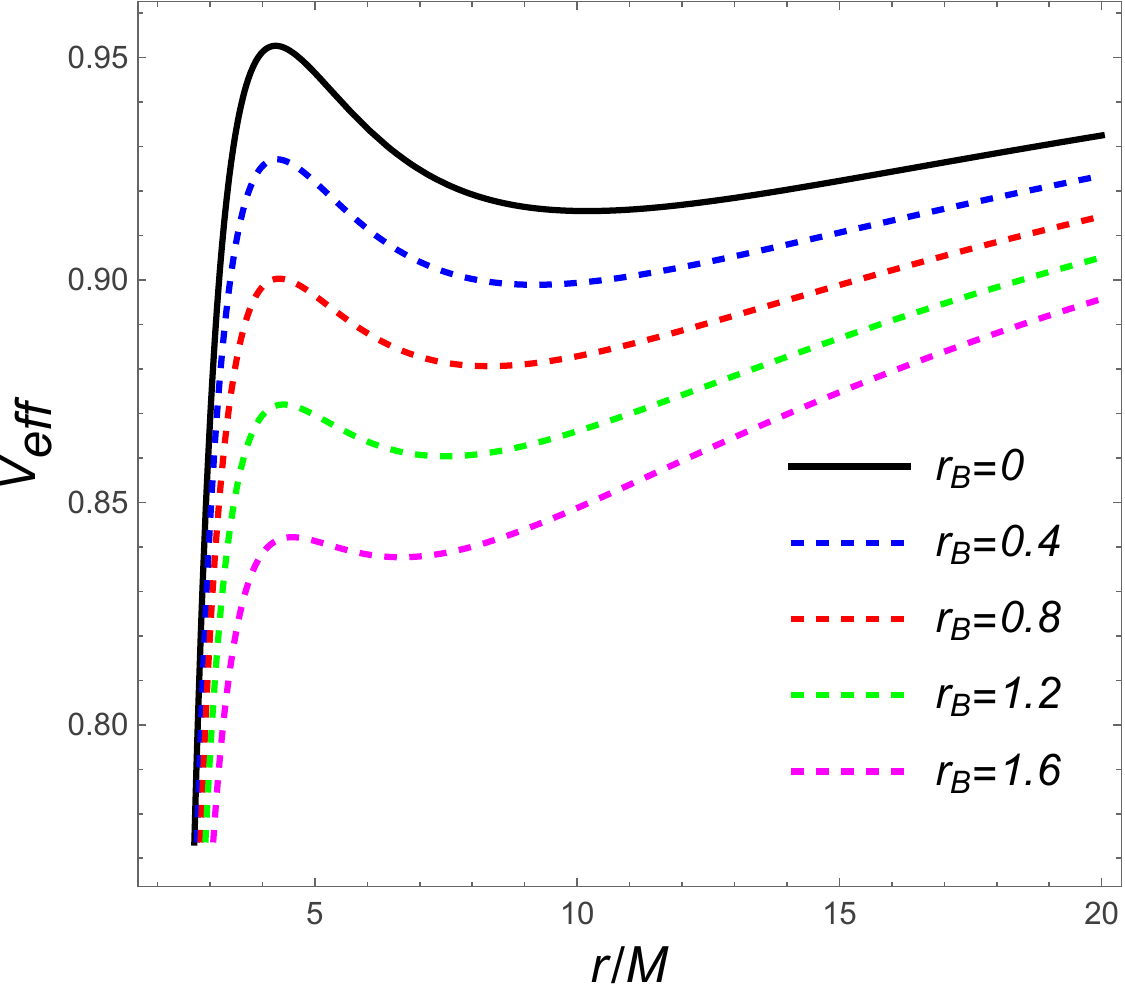}
		\caption{$L/M=3.8$}
		\label{Fig1b}
	\end{subfigure}
		\captionsetup{justification=raggedright,singlelinecheck=false} 
	\caption{Effective potential~\eqref{eq2} for test particle motion in charged black holes with scalar hair. (\ref{Fig1a}) Potential profiles for varying angular momentum at fixed $r_B$. (\ref{Fig1b}) Behavior for varying $r_B$ at constant angular momentum.}\label{Fig1}
\end{figure}

Figure~\ref{Fig1} shows the effective potential behavior. Figure~\ref{Fig1a} demonstrates that as orbital angular momentum increases, the potential develops extrema with minima and maxima corresponding to stable and unstable circular orbits, respectively. Figure~\ref{Fig1b} shows the potential maximum rises as $r_B$ decreases. As $r \to +\infty$, $V_{\text{eff}} \to 1$, implying $E=1$ serves as the critical energy separating bound from unbound motion, with bound trajectories satisfying $E \leq 1$.

The effective potential structure determines stable and unstable circular orbit locations, constraining allowed particle energies and angular momenta. For massive particles orbiting such black holes, bound trajectories exist between MBOs and ISCOs, with energy confined to
\begin{align}\label{eq5}
	E_{\text{ISCO}} \leq E \leq E_{\text{MBO}} = 1.
\end{align}
Here, $E_{\text{MBO}} = 1$ corresponds to particles having just enough energy to escape to infinity from the MBO, while $E_{\text{ISCO}}$ represents ISCO energy; particles with lower energies inevitably fall into the black hole.

Similarly, bound orbit angular momentum is restricted by
\begin{align}\label{eq6}
	L \geq L_{\text{ISCO}},
\end{align}
where $L_{\text{ISCO}}$ denotes ISCO angular momentum. These effective potential features define the allowed bound motion phase space around charged black holes with scalar hair.

\section{Marginally bound orbit and innermost stable circular orbit}\label{sec3}
We focus on MBOs and ISCOs for massive test particles around charged black holes. The MBO corresponds to the minimum-radius circular orbit allowing gravitational binding, with particles having just enough energy to escape to infinity. For the effective potential $V_{\rm eff}$, MBOs satisfy
\begin{align}\label{eq3}
	V_{\rm eff}(r_{\rm MBO}) &= 1,\\
	\frac{dV_{\rm eff}}{dr}\Big|_{r_{\rm MBO}} &= 0,
\end{align}
ensuring circular orbits without stability requirements.

The ISCO represents the smallest radius maintaining stable circular trajectories, determined by
\begin{align}\label{eq4}
	V_{\rm eff}(r_{\rm ISCO}) &= E^2, \\
	\frac{dV_{\rm eff}}{dr}\Big|_{r_{\rm ISCO}} &= 0, \\
	 \frac{d^2V_{\rm eff}}{dr^2}\Big|_{r_{\rm ISCO}} &= 0,
\end{align}
where particles with sub-ISCO energies plunge into the black hole. These characteristic orbits define the radial range for bound circular motion around charged black holes with scalar hair.

\begin{figure}
	\centering
	\begin{subfigure}{0.8\linewidth}
		\centering
		\includegraphics[width=\linewidth]{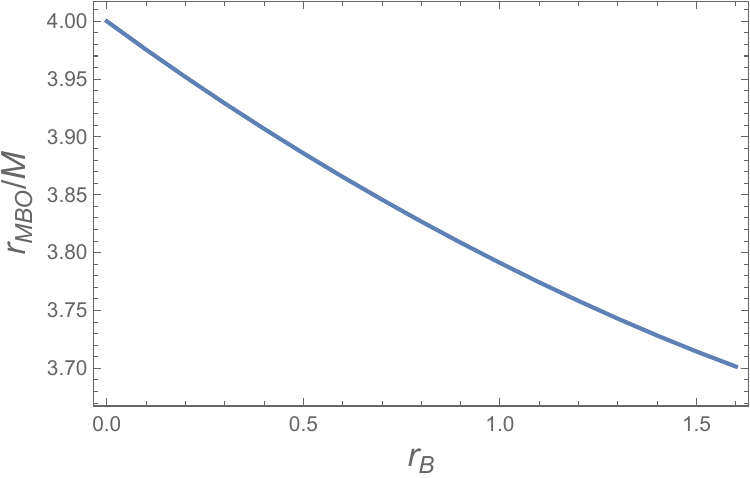}
		\caption{}
		\label{Fig2a}
	\end{subfigure}\\[6pt]
	\begin{subfigure}{0.8\linewidth}
		\centering
		\includegraphics[width=\linewidth]{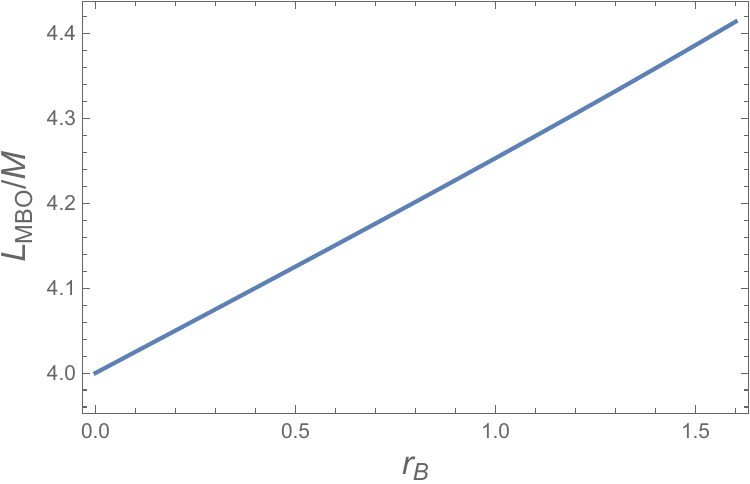}
		\caption{}
		\label{Fig2b}
	\end{subfigure}
		\captionsetup{justification=raggedright,singlelinecheck=false} 
	\caption{MBO properties around charged black holes with scalar hair. (a) MBO radius versus $r_B$. (b) MBO angular momentum versus $r_B$. Radius decreases while angular momentum increases with $r_B$.}
	\label{Fig2}
\end{figure}

To determine MBO characteristics, we employ Eq.~(\ref{eq3}). The resulting high-order algebraic equations prevent closed-form solutions, so we solve numerically. Figure~\ref{Fig2} shows MBO radius and angular momentum dependence on $r_B$. While MBO radius decreases with increasing $r_B$, the associated angular momentum grows monotonically, indicating scalar hair enables inward MBO shifts but requires larger angular momentum for marginal binding.

\begin{figure*}
	\centering
	\begin{subfigure}[b]{0.32\textwidth}
		\includegraphics[width=\textwidth]{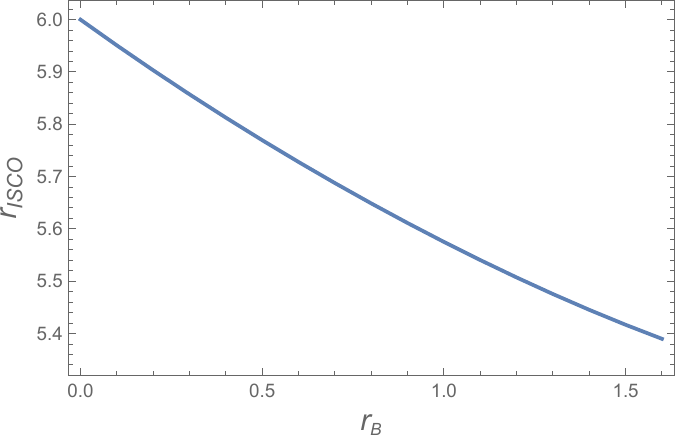}
		\caption{}
	\end{subfigure}
	\hspace{0.01\textwidth}
	\begin{subfigure}[b]{0.32\textwidth}
		\includegraphics[width=\textwidth]{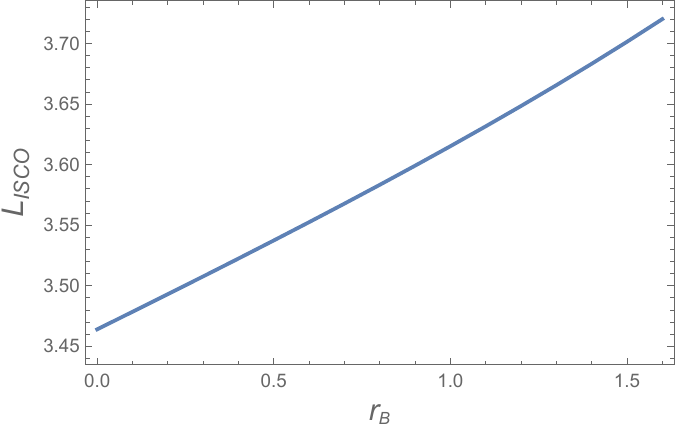}
		\caption{}
	\end{subfigure}
	\hspace{0.01\textwidth}
	\begin{subfigure}[b]{0.32\textwidth}
		\includegraphics[width=\textwidth]{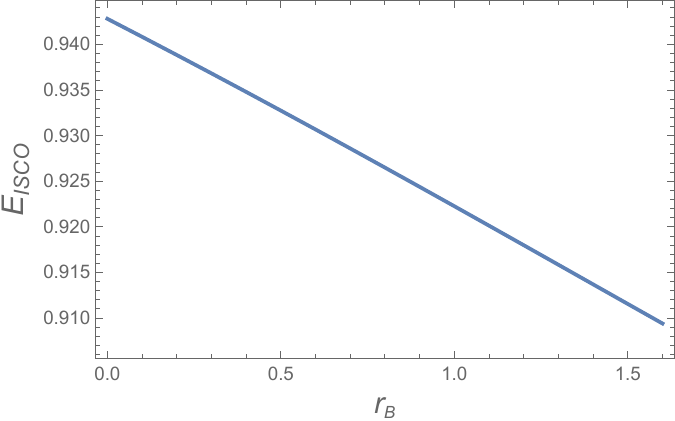}
		\caption{}
	\end{subfigure}
		\captionsetup{justification=raggedright,singlelinecheck=false} 
	\caption{ISCO properties around charged black holes with scalar hair. (a) ISCO radius versus $r_B$. (b) ISCO angular momentum versus $r_B$. (c) ISCO energy versus $r_B$. Radius and energy decrease while angular momentum increases with $r_B$.}\label{Fig3}
\end{figure*}

The ISCO, determined by Eq.~(\ref{eq4}), also depends on $r_B$. Numerical solutions yield ISCO radius, angular momentum, and energy variations with $r_B$, shown in Figure~\ref{Fig3}. ISCO radius and energy decrease with increasing $r_B$, while angular momentum increases, demonstrating scalar hair pushes ISCOs inward, reduces stability energy thresholds, but requires higher angular momentum for orbital stability.

\begin{figure}
	\centering
	\captionsetup{justification=raggedright,singlelinecheck=false} 
	\includegraphics[width=0.75\linewidth]{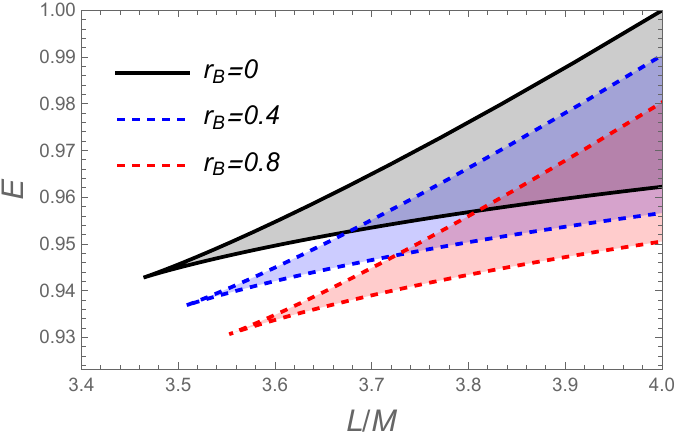}
	\caption{Allowed bound orbit parameter space around charged black holes with scalar hair, showing orbital angular momentum versus energy for different $r_B$ values.}
	\label{fig4}
\end{figure}

Finally, Figure~\ref{fig4} shows admissible bound orbit parameter space in the $(E,L)$ plane from Eqs.~(\ref{eq5}) and (\ref{eq6}). For fixed angular momentum, larger $r_B$ values extend the upper energy boundary, reflecting scalar hair's enlargement of energetically allowed bound orbits, enriching orbital dynamics.

\section{Periodic Orbits}\label{sec4}
\begin{figure}
	\centering
	\begin{subfigure}{0.7\linewidth}
		\centering
		\includegraphics[width=\linewidth]{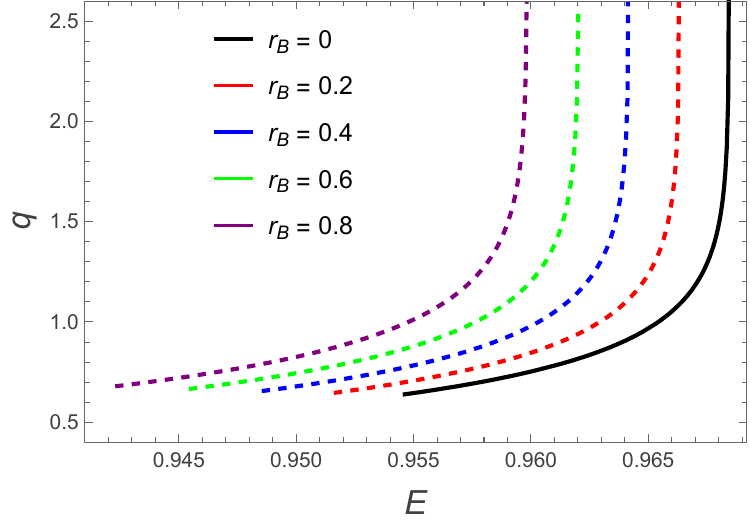}
		\caption{}
		\label{Fig5a}
	\end{subfigure}\\[6pt]
	\begin{subfigure}{0.7\linewidth}
		\centering
		\includegraphics[width=\linewidth]{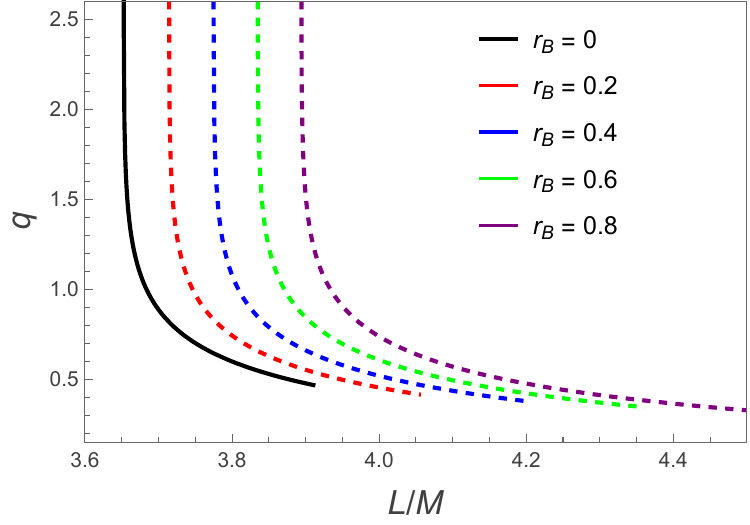}
		\caption{}
		\label{Fig5b}
	\end{subfigure}
		\captionsetup{justification=raggedright,singlelinecheck=false} 
	\caption{(a) Rational number $q$ versus periodic orbit energy for different $r_B$, with fixed $L=(L_{\text{MBO}}+L_{\text{ISCO}})/2$. (b) $q$ versus angular momentum for different $r_B$, with fixed $E=0.96$.}
	\label{Fig5}
\end{figure}

Having examined MBOs and ISCOs, we now consider periodic orbits near charged black holes with scalar hair. These occur when azimuthal-to-radial frequency ratios become rational. While generally $\omega_\varphi/\omega_r$ is irrational, it can be approximated by nearby rational values, enabling understanding of generic trajectories through neighboring periodic ones and their gravitational wave signatures \cite{Dy1}. Fundamental orbital frequencies are rationally related, with each orbit indexed by $(z,w,v)$ and rational parameter
\begin{align}\label{eqq}
	q \equiv \frac{\omega_\varphi}{\omega_r} - 1 = w + \frac{v}{z},
\end{align}
where $z$ (zoom), $w$ (whirl), and $v$ (vertex) are coprime integers. Using equations of motion (\ref{eqt})-(\ref{eqphi}) and (\ref{eqq}), this becomes
\begin{align}
	q = \frac{1}{\pi}\int_{r_2}^{r_1} \frac{L}{r^2\sqrt{f_B(r)}\sqrt{E^2 - V_\text{eff}(r)}}dr - 1,
\end{align}
where $r_1$ and $r_2$ denote periapsis and apoapsis, respectively. Fixing angular momentum as
\begin{align}
	L = \frac{L_{\text{MBO}} + L_{\text{ISCO}}}{2},
\end{align}
we plot $q$ versus orbital energy $E$ for different $r_B$ values. Similarly, fixing energy at $E=0.96$, we plot $q$ versus angular momentum $L$ for varying $r_B$, shown in Figure~\ref{Fig5}.

Figure~\ref{Fig5} shows $q$ increases slowly with $E$ initially but diverges as $E$ approaches its maximum; versus $L$, $q$ diverges near minimal $L$ then gradually decreases. For fixed $q$, orbital energy $E$ decreases while angular momentum $L$ increases with $r_B$; consequently, maximal orbital energy decreases with $r_B$ while minimal angular momentum increases. Although $r_B$ is physically expected small, we adopt relatively large values to qualitatively reveal its influence on periodic orbits and associated gravitational-wave signatures.

For periodic orbits indexed by different $(z,w,v)$ around charged black holes with varying $r_B$, we numerically calculate orbital energy $E$ with fixed $L=(L_{\text{MBO}}+L_{\text{ISCO}})/2$, and angular momentum $L$ with fixed $E=0.96$. Results appear in Tables~\ref{tab1} and~\ref{tab2}. Orbital energy decreases while angular momentum increases with $r_B$.

\begin{table*}[htbp]
	\centering
	\renewcommand{\arraystretch}{1.3} 
	\setlength{\tabcolsep}{8pt} 
	\begin{tabular}{ccccccccc}
		\hline
		\hline
		$r_B$ & $E_{(1,1,0)}$ & $E_{(1,2,0)}$ & $E_{(2,1,1)}$ & $E_{(2,2,1)}$ & $E_{(3,1,2)}$ & $E_{(3,2,2)}$ & $E_{(4,1,3)}$ & $E_{(4,2,3)}$ \\
		\hline
		0    & 0.96542525 & 0.96838267 & 0.96802639 & 0.96843422 & 0.96822477 & 0.96843840 & 0.96828487 & 0.96843968 \\
		\hline
		0.2  & 0.96293241 & 0.96625659 & 0.96584493 & 0.96631807 & 0.96607270 & 0.96632319 & 0.96614228 & 0.96632477 \\
		\hline
		0.4  & 0.96032030 & 0.96408470 & 0.96360387 & 0.96415914 & 0.96386802 & 0.96416553 & 0.96394946 & 0.96416751 \\
		\hline
		0.6  & 0.95763524 & 0.96192560 & 0.96135870 & 0.96201690 & 0.96166766 & 0.96202499 & 0.96176390 & 0.96202753 \\
		\hline
		0.8  & 0.95478014 & 0.95970156 & 0.95902643 & 0.95981519 & 0.95939109 & 0.95982561 & 0.95950598 & 0.95982893 \\
		\hline
		\hline
	\end{tabular}
	\captionsetup{singlelinecheck=off, justification=raggedright} 
	\caption{Energy $E$ for periodic orbits with different $(z, w, v)$ and $r_B$, with fixed $L=(L_{\text{MBO}} + L_{\text{ISCO}})/2$.}
	\label{tab1}
\end{table*}

\begin{table*}[htbp]
	\centering
	\renewcommand{\arraystretch}{1.3} 
	\setlength{\tabcolsep}{8pt} 
	\begin{tabular}{ccccccccc}
		\hline
		\hline
		$r_B$ & $L_{(1,1,0)}/M$ & $L_{(1,2,0)}/M$ & $L_{(2,1,1)}/M$ & $L_{(2,2,1)}/M$ & $L_{(3,1,2)}/M$ & $L_{(3,2,2)}/M$ & $L_{(4,1,3)}/M$ & $L_{(4,2,3)}/M$ \\
		\hline
		0    & 3.68358774 & 3.65340564 & 3.65759567 & 3.65270065 & 3.65533454 & 3.65263627 & 3.65462100 & 3.65261582 \\
		\hline				
		0.2  & 3.74626497 & 3.71506053 & 3.71933926 & 3.71435382 & 3.71702197 & 3.71429029 & 3.71629392 & 3.71427021 \\
		\hline		
		0.4  & 3.80857799 & 3.77564975 & 3.78016415 & 3.77490609 & 3.77771822 & 3.77483941 & 3.77695010 & 3.77481836 \\
		\hline		
		0.6  & 3.87092194 & 3.83556164 & 3.84045847 & 3.83474663 & 3.83781119 & 3.83467297 & 3.83697747 & 3.83464965 \\
		\hline		
		0.8  & 3.93368898 & 3.89514037 & 3.90058032 & 3.89421534 & 3.89765254 & 3.89413033 & 3.89672527 & 3.89410326 \\
		\hline		
		\hline
	\end{tabular}
	\captionsetup{singlelinecheck=off, justification=raggedright} 
	\caption{Angular momentum $L$ for periodic orbits with different $(z, w, v)$ and $r_B$, with fixed $E = 0.96$.}
	\label{tab2}
\end{table*}

\begin{figure*}
	\centering
	\includegraphics[width=\textwidth]{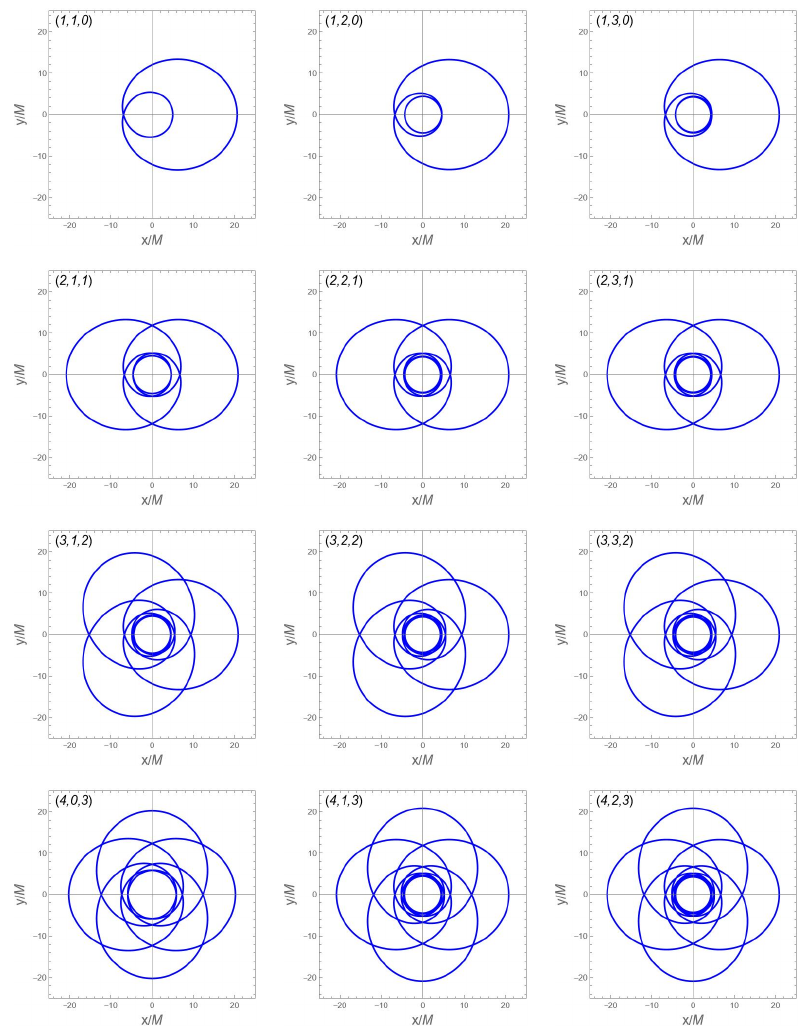}
		\captionsetup{justification=raggedright,singlelinecheck=false} 
	\caption{Periodic orbits with different $(z,w,v)$ around charged black holes with scalar hair, for $r_B=0.6$ and $E=0.96$.}\label{Fig6}
\end{figure*}

\begin{figure*}
	\centering
	\includegraphics[width=\textwidth]{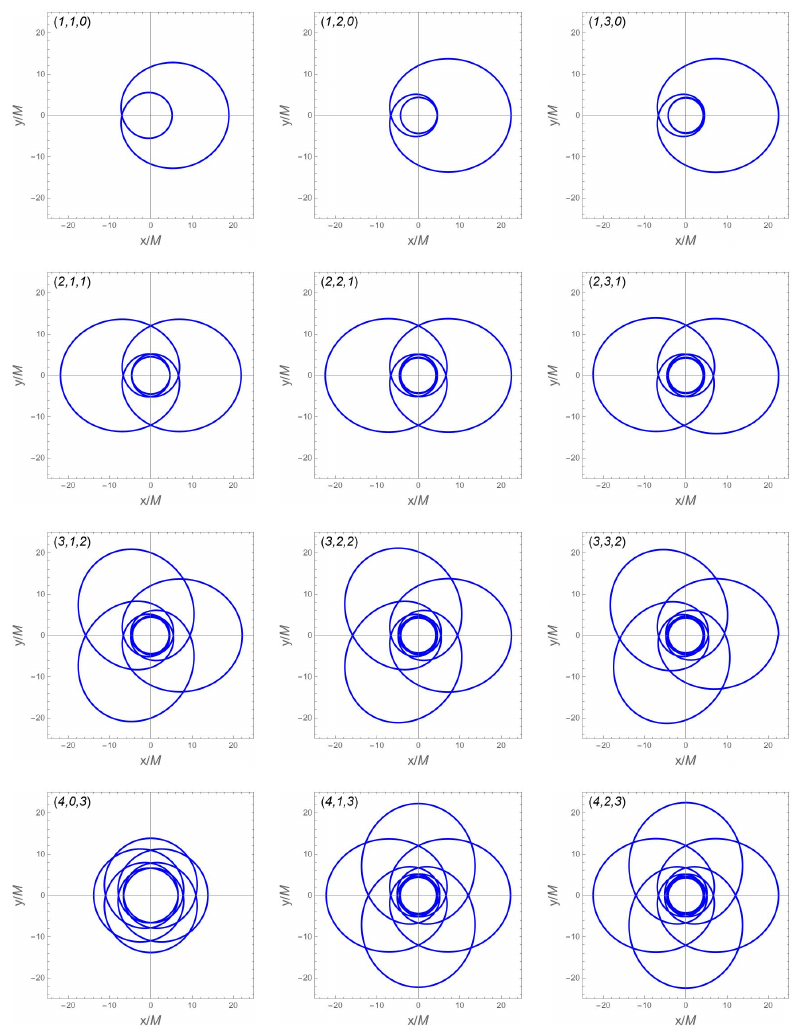}
		\captionsetup{justification=raggedright,singlelinecheck=false} 
	\caption{Periodic orbits with different $(z,w,v)$ around charged black holes with scalar hair, for $r_B=0.6$ and $L=\frac{(L_{\text{MBO}} + L_{\text{ISCO}})}{2}$.}\label{Fig7}
\end{figure*}

We also plot periodic orbits with different $(z,w,v)$ around charged black holes with $r_B=0.6$ in Figures~\ref{Fig6} and~\ref{Fig7}, with either fixed $E=0.96$ or $L=(L_{\rm MBO} + L_{\rm ISCO})/2$. Orbits with larger zoom numbers $z$ exhibit more intricate structures, while those with larger whirl numbers $w$ undergo more revolutions between successive apoapses.

\section{Numerical kludge gravitational waveforms from periodic orbits}\label{sec5}
A stellar-mass compact object on a periodic trajectory around a charged black hole with scalar hair constitutes an extreme mass-ratio inspiral (EMRI) system. Gravitational radiation from such systems potentially carries imprints of both periodic orbital structure and central black hole information. Standard EMRI waveform evaluation uses the adiabatic approximation \cite{A1,A2,A3,A4,poisson2014gravity,A5,Thorne:1980ru,A6,A7}. Since secondary orbital parameters evolve over timescales much longer than orbital periods, motion approximates geodesics with nearly constant energy and angular momentum over multiple cycles, neglecting radiation reaction in short intervals. We therefore compute gravitational waveforms during one complete orbital cycle.

We adopt the numerical kludge framework, effective for modeling EMRI waveforms \cite{klu}. Specifically, we numerically integrate equations of motion (\ref{eqmotion}) to obtain small body trajectories, then insert these into the quadrupole radiation formula \cite{Thorne:1980ru,poisson2014gravity} to generate waveforms. For a mass $m$ particle, the symmetric trace-free (STF) mass quadrupole moment is
\begin{align}\label{eqI}
	I^{ij} = \left[ \int d^3x\, x^ix^j T^{tt}(t,x^i) \right]_{\text{STF}},
\end{align}
with point particle stress-energy tensor on worldline $Z^i(t)$ given by
\begin{align}\label{eqT}
	T^{tt}(t,x^i) = m\,\delta^3(x^i - Z^i(t)).
\end{align}
Following Ref.~\cite{klu,P16}, we interpret Boyer-Lindquist coordinates as effective spherical polar coordinates, projecting trajectories to Cartesian form via
\begin{align}
	x &= r\sin\theta\cos\phi, \\
	y &= r\sin\theta\sin\phi, \\
	z &= r\cos\theta.
\end{align}
From Eqs.~(\ref{eqI})-(\ref{eqT}), metric perturbations describing gravitational-wave strain become
\begin{align}
	h_{ij} = \frac{2}{D_L}\frac{d^2I_{ij}}{dt^2}
	= \frac{2m}{D_L}(a_ix_j + a_jx_i + 2v_iv_j),
\end{align}
where $D_L$ is luminosity distance, and $v_i$ and $a_i$ are spatial velocity and acceleration. To relate perturbations to detector response, we introduce a detector-adapted frame $(X,Y,Z)$ \cite{poisson2014gravity}, centered on the supermassive black hole and aligned as
\begin{align}
	\mathbf{e}_X &= [\cos\zeta, -\sin\zeta, 0], \\
	\mathbf{e}_Y &= [\sin\iota\sin\zeta, -\cos\iota\cos\zeta, -\sin\iota], \\
	\mathbf{e}_Z &= [\sin\iota\sin\zeta, -\sin\iota\cos\zeta, \cos\iota],
\end{align}
where $\iota$ is orbital plane inclination relative to the $X$-$Y$ plane, and $\zeta$ is pericenter longitude. In this frame, the two gravitational-wave polarizations are
\begin{align}
	h_{+} &= \tfrac{1}{2}(e^i_Xe^j_X - e^i_Ye^j_Y)h_{ij}, \\
	h_{\times} &= \tfrac{1}{2}(e^i_Xe^j_Y + e^i_Ye^j_X)h_{ij},
\end{align}
yielding
\begin{align}
	h_{+} &= \tfrac{1}{2}(h_{\zeta\zeta} - h_{\iota\iota}), \\
	h_{\times} &= h_{\iota\zeta},
\end{align}
with components \cite{klu}
\begin{align}
	h_{\zeta\zeta} =& h_{xx}\cos^2\zeta - h_{xy}\sin 2\zeta + h_{yy}\sin^2\zeta, \\\nonumber
	h_{\iota\iota} =& \cos^2\iota\,[h_{xx}\sin^2\zeta + h_{xy}\sin 2\zeta + h_{yy}\cos^2\zeta] + h_{zz}\sin^2\iota \\\nonumber
	&- \sin 2\iota\,[h_{xz}\sin\zeta + h_{yz}\cos\zeta], \\
	h_{\iota\zeta} =& \cos\iota\left(\tfrac{1}{2}h_{xx}\sin 2\zeta + h_{xy}\cos 2\zeta - \tfrac{1}{2}h_{yy}\sin 2\zeta\right) \\\nonumber
	&+ \sin\iota\,[h_{yz}\sin\zeta - h_{xz}\cos\zeta].
\end{align}

\begin{figure*}[htbp]
	\centering
	\begin{subfigure}{\textwidth}
		\centering
		\includegraphics[width=\linewidth]{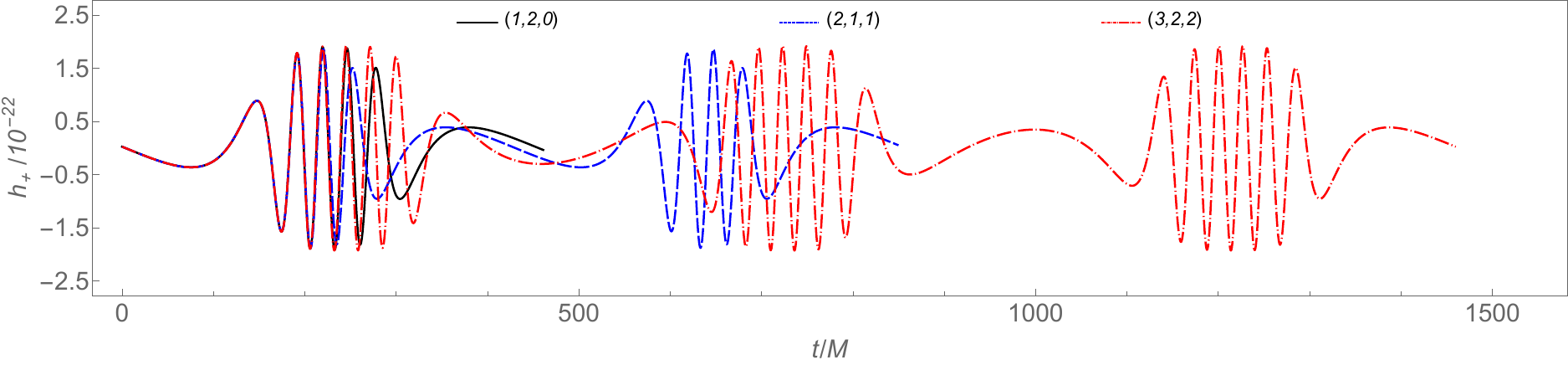}
		\caption{}
		\label{Fig8a}
	\end{subfigure}\\[6pt]
	\begin{subfigure}{\textwidth}
		\centering
		\includegraphics[width=\linewidth]{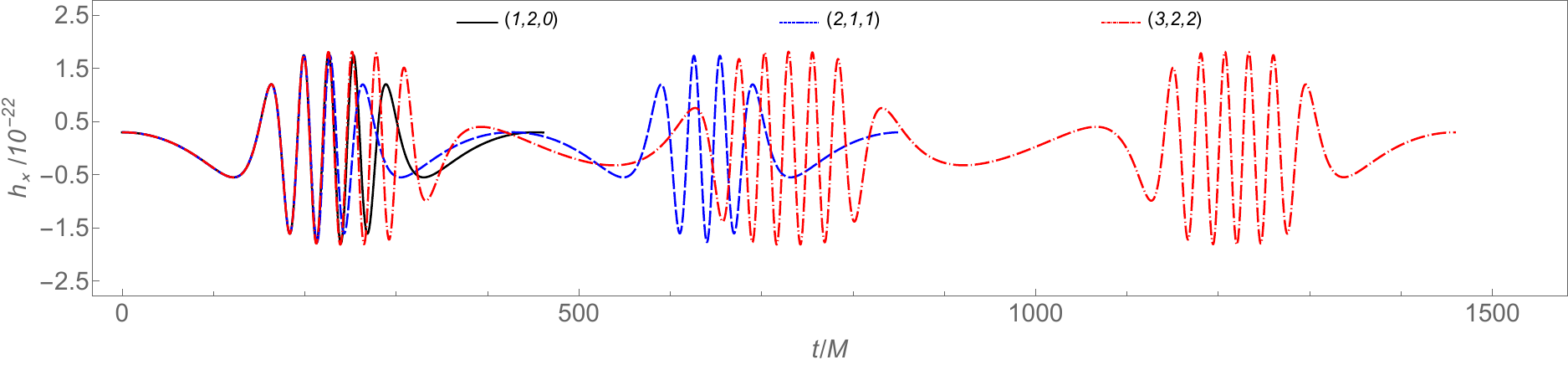}
		\caption{}
		\label{Fig8b}
	\end{subfigure}
		\captionsetup{justification=raggedright,singlelinecheck=false} 
	\caption{Gravitational waveforms from $m=10M_\odot$ test objects on different periodic orbits around $M=10^6M_\odot$ supermassive charged black holes with scalar hair, with fixed $E=0.96$.}
	\label{Fig8}
\end{figure*}

As a concrete example, we consider an EMRI with $m=10M_\odot$ stellar-mass object orbiting a $M=10^6M_\odot$ charged black hole with scalar hair at $D_L=2$ Gpc. For $\iota=\pi/4$, $\zeta=\pi/4$, and $r_B=0.6$, the corresponding polarizations appear in Figure~\ref{Fig8}. Waveforms show alternating zoom and whirl phases within single orbits, reflecting underlying periodic trajectory dynamics. Orbits with higher zoom indices $z$ produce more intricate waveform substructures, corresponding to multi-leaf orbital morphology.

\begin{figure*}[htbp]
	\centering
	\begin{subfigure}{\textwidth}
		\centering
		\includegraphics[width=\linewidth]{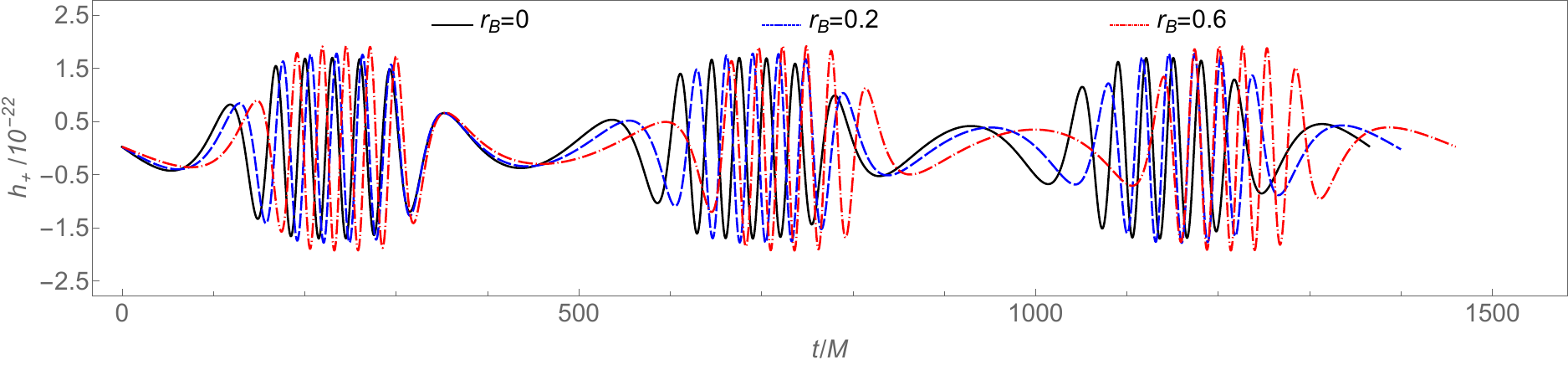}
		\caption{}
		\label{Fig9a}
	\end{subfigure}\\[6pt]
	\begin{subfigure}{\textwidth}
		\centering
		\includegraphics[width=\linewidth]{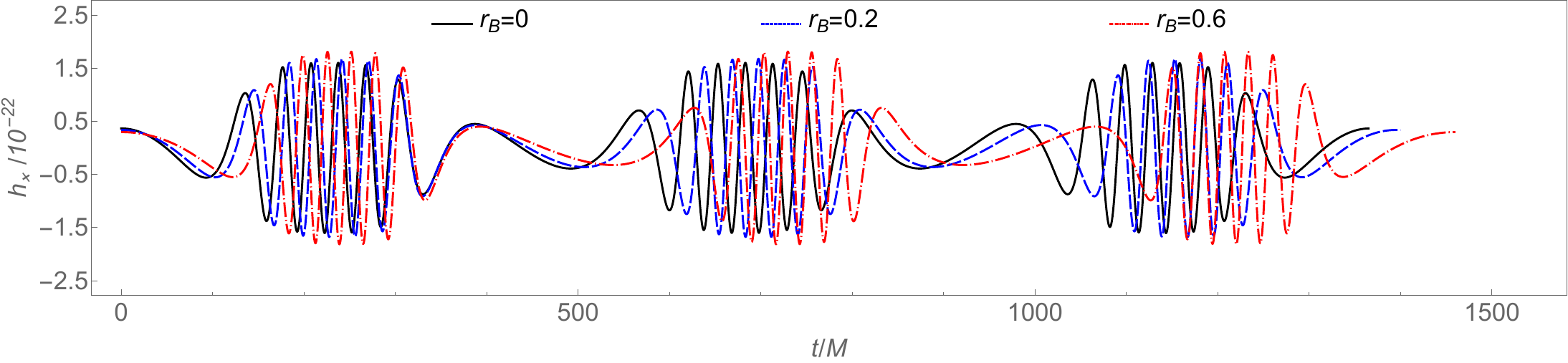}
		\caption{}
		\label{Fig9b}
	\end{subfigure}
		\captionsetup{justification=raggedright,singlelinecheck=false} 
	\caption{Gravitational waveforms from $m=10M_\odot$ test objects on $(3,2,2)$ periodic orbits around $M=10^6M_\odot$ supermassive charged black holes with scalar hair, for different $r_B$, with fixed $E=0.96$.}
	\label{Fig9}
\end{figure*}

Under identical setup, Figure~\ref{Fig9} compares waveforms from fixed $(3,2,2)$ periodic orbits for varying $r_B$. Increasing $r_B$ modestly modifies waveform amplitude while inducing cumulative phase advancement. This phase shift builds gradually over time, becoming increasingly significant during orbital evolution.

\section{Conclusions and outlook}\label{sec6}
This work systematically investigated geodesic motion and associated gravitational wave signatures in charged black hole spacetimes with scalar hair. Using the effective potential approach, we analyzed marginally bound orbits (MBOs) and innermost stable circular orbits (ISCOs). Results indicate the scalar hair parameter $r_B$ plays a significant role: increasing $r_B$ shifts both MBOs and ISCOs inward, decreases corresponding orbital energies, and increases angular momenta required for stable motion. Consequently, the allowed bound orbit range in $(E,L)$ parameter space effectively extends.

We further conducted detailed studies of periodic orbits characterized by rational winding numbers and classified by integer indices $(z,w,v)$. Analysis shows increasing $r_B$ generally reduces energies and raises angular momenta across different $(z,w,v)$ combinations, highlighting scalar hair's universal influence on orbital stability and underlying dynamical structure. Moreover, orbits with larger zoom or whirl numbers exhibit increasingly intricate geometrical features, including multiple near-horizon revolutions followed by rapid excursions, forming characteristic zoom-whirl behavior. These findings demonstrate scalar hair not only modifies periodic orbit distribution in energy-angular momentum space but also significantly impacts morphological characteristics, revealing rich nonlinear dynamics in such spacetimes.

Building on this, we simulated gravitational waveforms from extreme mass-ratio inspirals (EMRIs) where stellar-mass compact objects move along periodic orbits around supermassive charged black holes with scalar hair. Numerical approximations reveal clear zoom-whirl signatures in waveforms, directly corresponding to underlying orbital geometry. Notably, $r_B$ variations induce observable modifications in both waveform amplitude and phase, with cumulative phase shifts becoming increasingly significant during inspiral. These results underscore gravitational-wave astronomy's potential—particularly future space-based detectors like LISA—to probe deviations from Kerr black holes and constrain scalar hair presence in astrophysical black holes.

Looking ahead, several avenues merit investigation. Incorporating radiation reaction and full self-force corrections would enable more precise long-term EMRI evolution modeling. Extending analysis to rotating or non-equatorial orbits could further enrich waveform phenomenology. Finally, confronting predictions with actual detector sensitivity will be essential for assessing feasibility of testing alternative black hole models and potential extra-dimensional effects through gravitational-wave observations.
    
\acknowledgements

This work was supported by the National Natural Science Foundation of China (Grants No. 12035005, No. 12405055, and No. 12347111), the China Postdoctoral Science Foundation (Grant No. 2023M741148), the Postdoctoral Fellowship Program of CPSF (Grant No. GZC20240458), and the National Key Research and Development Program of China (Grant No. 2020YFC2201400).

\bibliography{Refs}

\end{document}